# π-molecular dielectric layer for organic thin film diode

*Céline Trapes, Virginie Gadenne(*), Lamia Rouaï*

LACSC, ECE, 53 rue de Grenelle, 75007 Paris
(*) L2MP-UMR CNRS 6137, ISEN, Place George Pompidou, 83000 Toulon

**ABSTRACT**

Very thin (1.2-2.5nm) self-assembled organic dielectric monolayers have been integrated into organic thin-film diode to achieve electrical characteristics. These dielectrics are fabricated by self-assembling deposition, resulting in smooth, strongly adherent, thermally stable, organosiloxane thin films having interesting electrical capacitances (around 150 $nF.cm^{-2}$ at -3V) and insulating properties (leakage current densities around $10^{-5}$ $A.cm^2$ at -1V).

## 1. INTRODUCTION

Organic thin-film transistors (OTFTs) based on π-electron materials are envisioned as critical components of future organics technologies and would enable low-cost solution-processed/ printed logic circuits, sensors [1-3].
This work is an investigation using self-assembled molecules (SAMs) to develop organic insulators [2]. (100) silicon substrate $10^{-15}cm^{-3}$ p-type doped is modified by grafting organic molecules onto its surface by using wet chemistry methods. Previous studies have been achieved with aliphatic insulating molecules as octadecyltrichlorosilane (C18) in the laboratory [7]. This time we are studying molecules rich in **π** electron as phenylbutyltrichlorosilane (PBTCL) and pentafluorophenyltrichlorosilane (FPPTCL) purchased from ABCR GmbH and Co. Also in order to elaborate mix layers, propyltrichlorosilane (C3) are used. Formation of self-assembled monolayers (SAMs) is controled using ellipsometry to determine the thickness, goniometry and atomic force microscopy (AFM) to analyse the surface topography. Metal-SAM-Si samples are tested as resistive diode using current-voltage or capacity-voltage measurements.

## 2. MAIN RESULTS

*Goniometry*

Thanks to Krüss DSA10-MK2 in ambient atmosphere, contact angles are measured with a deionised water drop (~5µl) with a precision of about one degree. Our results correspond to literature [4-5] where typical contact angle of phenyl end group (PBTCl) is around 90° and around 110° for alkyl chain (C3 or C18) with a measurement error-bar around ±2° for 10 measurements per sample.

*Ellipsometer*

Ellipsometer used was a Scentech SE400 with Helium-neon beam of 632.8 nm forming a 70° angle with the device surface. Taking a refractive index of 3.865 for silicon substrate, 1.46 for native oxide, and 1.5 for aromatic monolayer [6], we have extracted thicknesses of 20 Å for $SiO_2$ 12 Å for PPTCL or FPPTCL and 25 Å for C18 with an error bar of ±3 Å .

*Electrical Measurements I-V*

I-V measurements are performed thanks to an HP 4155 current analyser and a digital capacitance meter in a [-3V; 2V] voltage range.
For each I-V measurement, we extract the flat-band voltage ($V_{FB}$), and the conductivity in the ohmic range defined as [$-0.2V+V_{FB}$;$+0.2V+V_{FB}$]. Then we extract the conductivity thanks to the current density J (V) as $\sigma = e\left(dJ/dV\right)$ where e is the SAM thickness.
Positive measurements are classically saturating as the bulk is lightly doped ($p=10^{-15}cm^{-3}$) (figure 1.a) [9]. I-V measurement in accumulation regime is diminishing (figure 1.b) relatively to the insulating properties of each SAM tested.
But one question should be pointed out: π end group rich in electron of PBTCl and FPPTCl should not improve the conductivity effect? As conductivities extracted σ are about $10^{-12}$ $S.cm^{-1}$ for $SiO_2$, $7.510^{-14}$ $S.cm^{-1}$ for FPPTCl and 2.5 $10^{-14}$ $S.cm^{-1}$ for PBTCl, this effect is not verified here.
This could be explained with the relative augmentation of the SAM thickness: in comparison with native oxide diode, the addition of PBTCl or FPPTCl doubles the total SAM thickness. As a result, the thickness SAM augmentation hides the conducting effect of π end group.

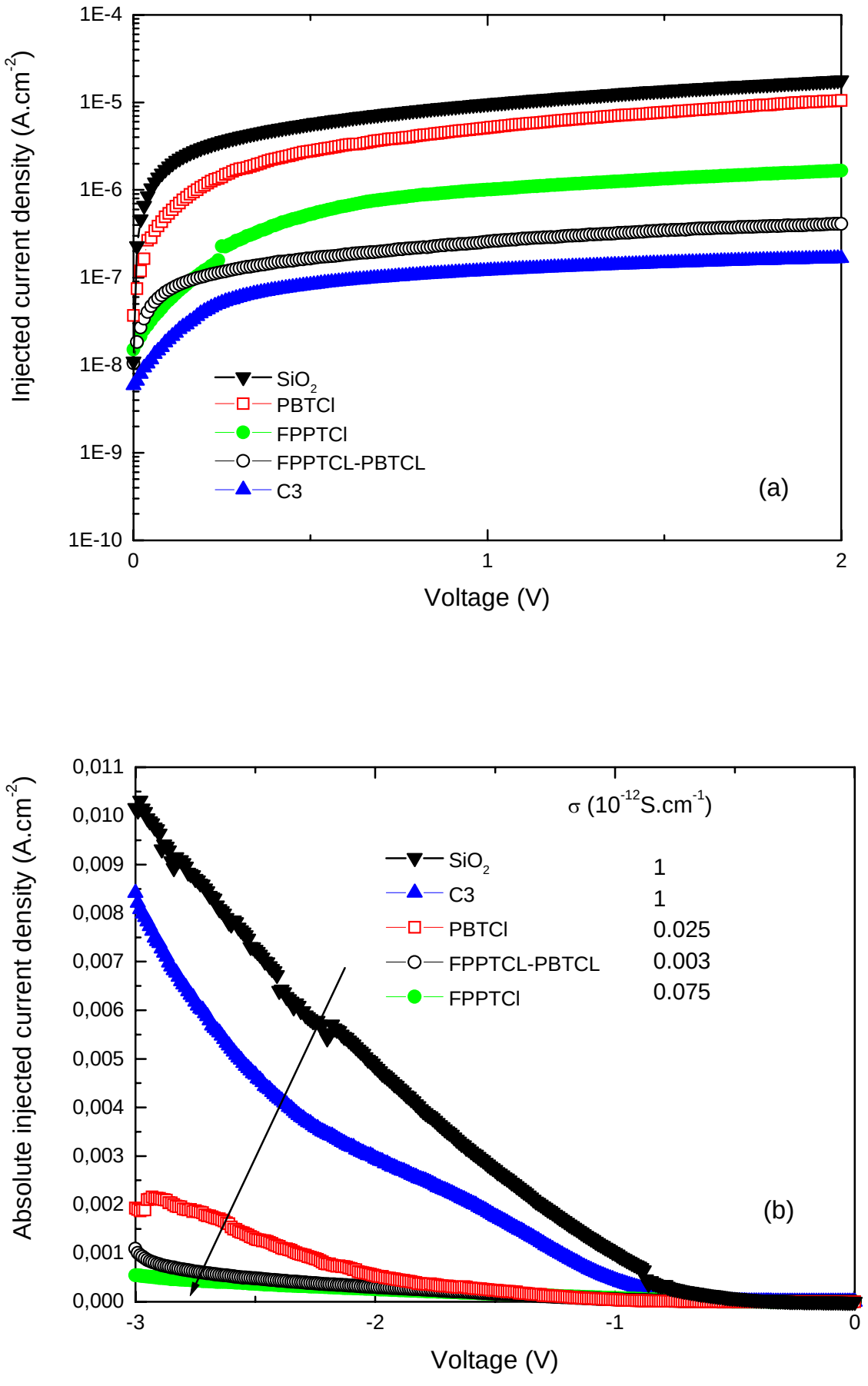

Fig. 1 (a) and (b): Absolute current density vs volage (positive range (a) and accumulation regime focus (b)).

*Electrical Measurements C-V*

C-V characteristics are obtained for a signal of 800kHz of frequency and 15 mV of amplitude in addition with a constant one varying from -3V to 0.5V.

In reference to MOS (Metal-Oxide-Semiconductor) C-V macro-model [8], we could use one in accumulation regime ($V_G$<0) as a resistance $R_S$ (parasitic effects), and the serial composition of two capacitances $C_X$ and $C_{SiO2}$. Considering this assumption, the total measured capacitance $C_{MAX}$ can be considered as:

$$C_{MAX} = \left(1/C_X + 1/C_{SIO2}\right)^{-1} \text{ and } C_X = \frac{\varepsilon_0 \varepsilon_x S}{T_x} \quad [1]$$

Where x is the SAM to characterise, $\varepsilon_x$ is the dielectric constant, $T_x$ the monolayer thickness, and $\varepsilon_o$ the vacuum permittivity.

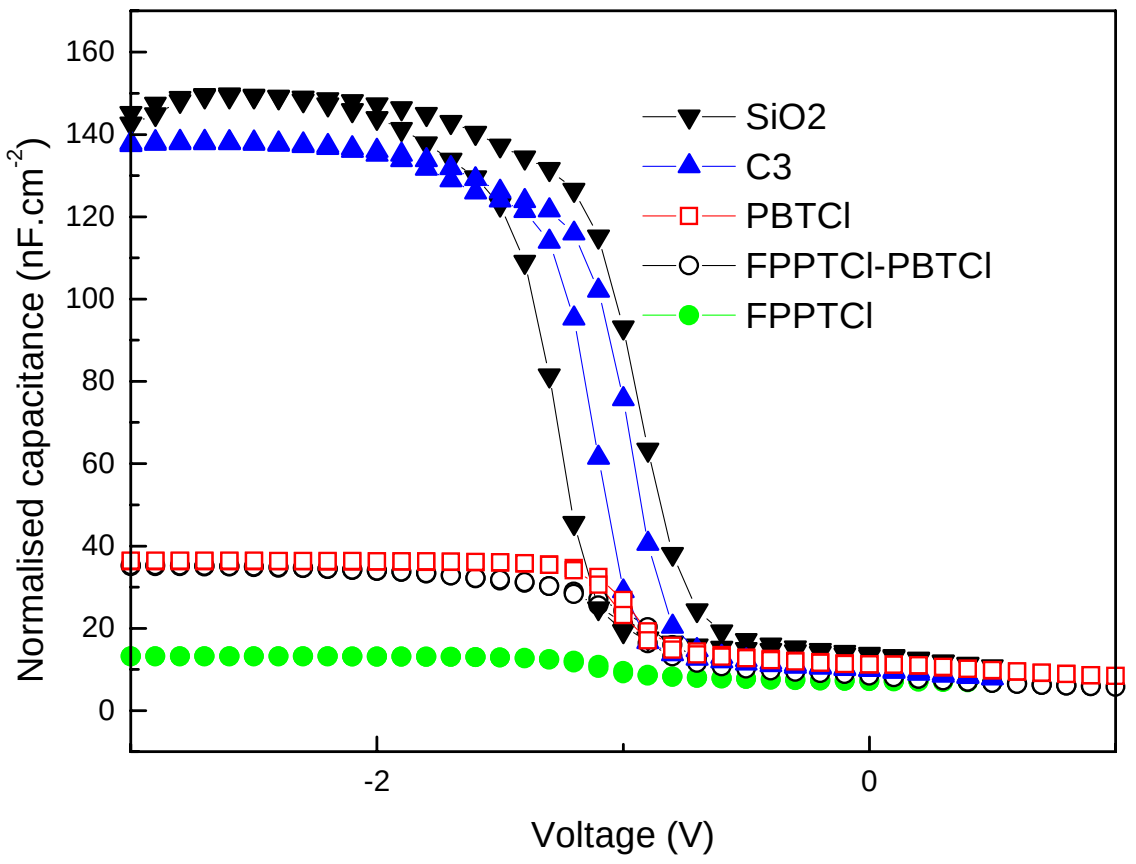

Fig. 2 : 800kHz C-V characteristics – sweep rate 5mV/s of SAM diode

An hysteresis of 0.1V for C3 and 0.3V for $SiO_2$ is observed with a shift towards the negative gate voltage. This indicates the presence of positive charges ($N_{OT}$) in the insulator, and we estimated their concentration to be about $10^{11}$cm$^{-2}$ for C3 or $SiO_2$. This effect is lighter with SAM like PBTCl, FPPTCL or mix PBTCL-FPPTCL measurement ($N_{OT}$~5 $10^{10}$ cm$^{-2}$).

In accumulation regime, the maximum value is diminishing with PBTCL and FPPTCL additional SAM. This effect could be explained in two ways: the first considering the SAM thickness augmentation, which is correlated with the diminution of the conductivity. Another explanation would be that the aromatic end group would be a barrier to the metal diffusion in comparison with native oxide SAM.

But these values are too low in comparison with theoretical values (with a ratio of ten) and we can not extract SAM permittivity from the C-V measurement with the classical model considered below (formula 1).

## 3. PERSPECTIVES

Even if the conductivity is significantly reduced with aromatic SAM, which has been traduced as reduced leakage current throw the monolayer, electrical measurement do not already allow characterising the SAM structure. This study should be follow in improving metal deposition (controlled atmosphere, and slower method deposition in order to limit the damages) or extracting a new macro-model considering aluminium oxide interface.

## 4. ACKNOWLEDGMENT

Many thanks to the "Laboratoire Matériaux Microélectronique de Provence" (L2MP) for their supports in samples preparation and characterization.

## 5. REFERENCES

[1] C.Kagan, C.R. Andry, and E.F. Roberts, “Thin film transistors,” *cdsl*, Marcel Dekker, New-York, pp. 377-425, 2003.

[2] C.D Dimitrakopoulod, P.R.L. Malenfant, *Adv. Mater.*,**14**, pp. 99-117 Publisher, 2002.

[3] H. Sirringhaus, *Nat. Mat.*,**2**, pp. 641-642, 2002.

[4] M. W. Tsao, C.L Hoffmann and J. F. Rabolt, *Langmuir.*,**14**, pp. 4317-4322, 1997.

[5] C. Ming Yam, S. S. Y Tong and A. K. Kakkar, *Langmuir.*,**14**, pp. 6941-6947, 1998.

[6] S. Lenfant, phd, Flandre-Artois University, 2001.

[7] S. Desbief, phd, Provence University, 2006.

[8] P. Fontaine, D. Goguenheim, D. Deresmes, D. Vuillaume, *Appl. Phys. Lette*r, phd, pp. 2256-2258, 2001.

[9] H. Mathieu, “Physique des semiconducteurs et des composants électroniques”, 2001.